\documentclass[superscriptaddress,twocolumn, secnumarabic, amssymb,amsmath,amsfonts,floatfix,longbibliography,aps,prl,10pt]{revtex4-2}

\usepackage{chemformula,siunitx}
\usepackage{lastpage}
\usepackage[protrusion=true, expansion=true]{microtype}
\usepackage{wrapfig}
\usepackage{multirow}
\usepackage{array}
\usepackage{ragged2e}
\justifying

\usepackage{graphicx}
\usepackage{epstopdf}
\usepackage[T1]{fontenc}
\usepackage{amsbsy}
\usepackage[T1]{fontenc}
\usepackage{amsmath}
\usepackage{amssymb}
\usepackage{bbm}
\usepackage{braket}
\usepackage{xcolor}
\allowdisplaybreaks
\usepackage{graphicx}
\usepackage{booktabs}
\usepackage{makecell}
\usepackage[normalem]{ulem}

\newcommand{\equref}[1]{Eq.~(\ref{#1})}

\newcommand{\figref}[1]{Fig.~\ref{#1}}

\newcommand{\tableref}[1]{Table~\ref{#1}}

\usepackage[colorlinks=true]{hyperref}
\hypersetup{
    unicode=false,          
    pdftoolbar=true,        
    pdfmenubar=true,        
    pdffitwindow=false,     
    pdfstartview={FitH},    
    pdftitle={Ta1-xWxSe2},    
    pdfauthor={Poulami Manna},      
    pdfsubject={},   
    pdfcreator={},   
    pdfproducer={}, 
    pdfkeywords={} {} {}, 
    pdfnewwindow=true,      
    colorlinks=true,       
    linkcolor=blue, 
    citecolor=blue,        
    filecolor=magenta,      
    urlcolor=blue          
}

\begin{document}

\title{\textrm{Emergence of Quasi-two-dimensional Superconductivity in W-doped Bulk Noncentrosymmetric 3$R$-TaSe$_2$}}

\author{P. Manna}
\affiliation{Department of Physics, Indian Institute of Science Education and Research Bhopal, Bhopal, 462066, India}
\author{R.~P.~Singh}
\email[]{rpsingh@iiserb.ac.in}
\affiliation{Department of Physics, Indian Institute of Science Education and Research Bhopal, Bhopal, 462066, India}

\begin{abstract}
Noncentrosymmetric transition-metal dichalcogenides offer a rich environment for the study of unconventional superconducting phenomena. Here, we present a comprehensive analysis of single-crystalline W-doped 3$R$-TaSe$_2$, revealing weakly coupled anisotropic unconventional superconductivity at $T_c$ = 2.82(2) K, with an in-plane upper critical field exceeding the Pauli limit by 1.7 times. The angular dependence of the upper critical field, along with the observation of a Berezinskii-Kosterlitz-Thouless transition, reveals quasi-two-dimensional superconductivity. Crucially, magnetotransport reveals a distinct two-fold rotational symmetry within the superconducting state under in-plane fields, breaking the underlying three-fold lattice symmetry. These findings establish W-doped $3R\text{-TaSe}_2$ as a bulk model system for exploring intrinsic low-dimensional superconductivity and broken rotational symmetry, thus opening new directions for future quantum technologies.
\end{abstract}

\maketitle

\section{INTRODUCTION} 
Reducing dimensionality from bulk to monolayer limit in two-dimensional (2D) superconductors unlocks a broad spectrum of novel quantum states, including the Berezinskii-Kosterlitz-Thouless (BKT) transition, Ising pairing, the charge density wave (CDW) transition, and field-induced mixed parity states \cite{lian2023interplay, qiu2021recent, PhysRevB.100.064506, xi2016ising, m9xx-gk46, hamill2021two}. In noncentrosymmetric systems, broken inversion symmetry-driven antisymmetric spin-orbit coupling (ASOC) stabilizes Ising superconductivity with spins locked out-of plane, leading to in-plane upper critical field exceeding the Pauli limit \cite{de2018tuning}. In addition to Ising superconductivity, spin-orbit coupling (SOC) can also cause field-induced mixed-parity pairing \cite{PhysRevB.99.180505} and topological superconductivity \cite{PhysRevLett.113.097001}. Thus, reduced dimensionality and inversion symmetry breaking act together, intrinsically intertwining various unconventional pairing and 2D superconductivity \cite{https://doi.org/10.1002/adma.202312341}. Transition-metal dichalcogenides (TMDs) provide a versatile platform for exploring such physics, where chemical doping, intercalation, and heterostructure engineering tune low-dimensional superconductivity. Notably, recent studies show that 2D superconducting properties can also be achieved in bulk systems \cite{devarakonda2020clean} with sufficiently weakened interlayer coupling. This greatly extends the range of experimentally accessible systems beyond monolayers \cite{PhysRevMaterials.4.124803, lu2015evidence, PhysRevB.106.134515, agarwal2023quasi, shi2024two, fan2025two, gmcz-mw9g}. In this context, chemical doping enables precise control over the density of states near the Fermi surface, while maintaining structural stability \cite{Hu_2021}.

Understanding unconventional superconductivity in low-dimensional systems remains a central challenge in condensed matter physics. 5d-TMDs offer a particularly fertile ground in this regard, as their strong intrinsic spin–orbit coupling (SOC) can stabilize exotic pairing states. Among these, TaS$_2$ has emerged as a model system, with its diverse polymorphs hosting a remarkable variety of unconventional superconducting phenomena—including Ising pairing, chiral superconductivity, and nematic superconductivity \cite{d13p-mtbz, liu2024nematic, 6rc6-mm3b, doi:10.1126/sciadv.aax9480}. However, its isostructural counterpart TaSe$_2$ remains comparatively underexplored. Its 2$H$ polymorph, consisting of a centrosymmetric crystal structure, implicitly restricts the impact of ASOC. In contrast, noncentrosymmetric 3$R$-TaSe$_2$ provides an ideal platform to investigate the interplay of reduced dimensionality, strong SOC, and broken inversion symmetry \cite{https://doi.org/10.1002/adfm.202501453}.
However, while the 2$H$ variant is the most stable, the synthesis and stabilization of the 3$R$ phase remain the subject of ongoing debate \cite{luo2015superconductivity, Deng2020}. Intriguingly, despite this challenge, a recent work revealed that 3$R$-TaSe$_2$ exhibits a markedly increased superconducting transition at 2.89 K, compared to 2$H$-TaSe$_2$, which shows a very low superconducting transition temperature (0.14 K) \cite{bhoi2016interplay}. Moreover, the presence of the CDW state near 114 K along with superconductivity in 3$R$-TaSe$_2$ suggests a remarkable electronic ground state \cite{https://doi.org/10.1002/adfm.202501453, dharmasiri2026charge}. 

Chemical tuning offers a powerful route to further manipulate the superconducting properties of TMDs. Self-intercalation \cite{Bai_2018}, as well as intercalation and substitution with Cu, Pd, Pt, Te \cite{luo2015superconductivity, bhoi2016interplay, PhysRevB.107.104510, doi:10.1073/pnas.1502460112, qi2026charge}, have been shown to enhance T$_c$ and alter stacking sequences, while Sn intercalation in 3$R$-TaSe$_2$  has revealed three-dimensional Ising superconductivity \cite{doi:10.1021/acs.nanolett.5c00196}. Collectively, these results indicate that the superconducting properties of TaSe$_2$ are highly sensitive to interlayer coupling. This raises the question, can doping-induced weakening of interlayer coupling stabilize quasi-2D superconductivity in bulk TaSe$_2$. Recent results on doped 2$H$-TaS$_2$ and 2$H$-NbSe$_2$ \cite{PhysRevB.106.134515, patra2024planar} support this scenario, suggesting that analogous doping in TaSe$_2$ may weaken interlayer coupling, induce quasi-2D superconductivity and unconventional pairing. Since achieving two-dimensional superconductivity conventionally requires monolayer isolation, realizing it instead through controlled interlayer decoupling in bulk crystals offers a more accessible route to low-dimensional physics and simpler device architectures.

\begin{figure*}
\includegraphics[width=1.99\columnwidth]{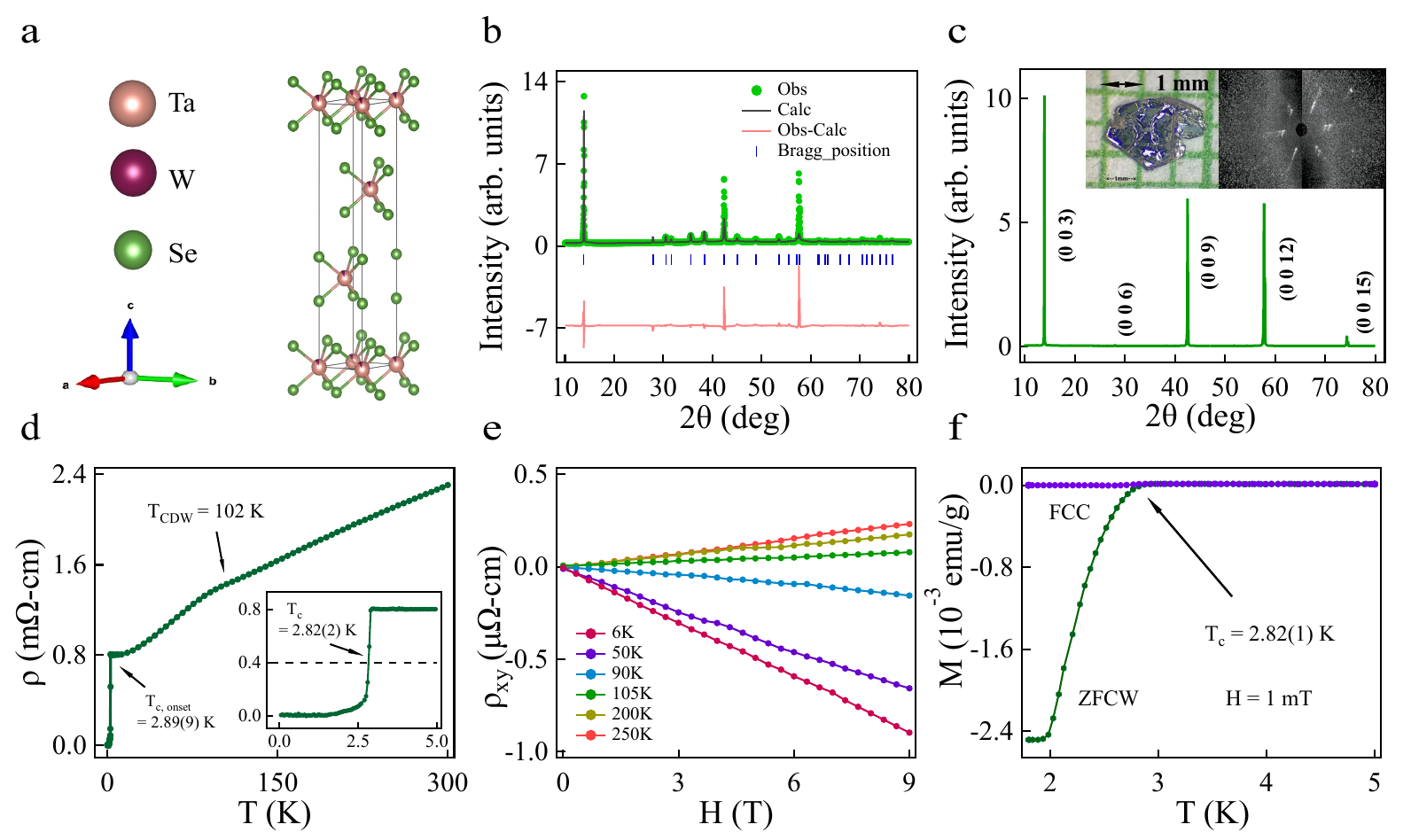}
\caption {\label{fig1}(a) The crystal structure of 10 $\%$ W-doped TaSe$_2$, where Ta, W, Se atoms are represented by peach, wine red, and green colors, respectively.  (b) Powder XRD patterns of crushed crystal are refined with the Rietveld method using the 3R-TaSe$_2$ structure. (c) Single-crystal XRD patterns oriented along the $c$-axis with (00$n$) reflections. Left inset: microscopic image of a 3 mm long crystal. Right inset: Laue diffraction pattern. (d) Zero-field electrical resistivity is plotted as a function of temperature from 0.05 K to 300 K, with the inset showing the superconducting transition at $T_c$ = 2.82(2) K. The anomaly around 102 K indicates a CDW transition. (e) The Hall resistivity varies with temperature, reflecting changes in carrier type on either side of the CDW transition. (f) The temperature-dependent magnetization shows superconductivity at 2.82(1) K.}
\end{figure*}

In our study, we address these challenges by successfully synthesizing crystalline 3$R$-Ta$_{1-x}$W$_x$Se$_2$ for $x$ = 0.1 and report detailed investigations of its superconducting properties through magnetization, specific heat, and transport measurements. The results confirm weakly coupled anisotropic type-II superconductivity in bulk crystals, with a transition temperature of $T_c$ = 2.82(2) K, with in-plane upper critical field exceeding the Pauli limit. Angle-dependent magnetotransport measurements and clear signatures of a BKT transition provide strong evidence for quasi-2D superconductivity. Moreover, the superconducting state exhibits rotational symmetry breaking in the presence of in-plane magnetic fields, thereby highlighting an unusual superconducting state in this compound.

\begin{figure*}
\includegraphics[width=1.98\columnwidth]{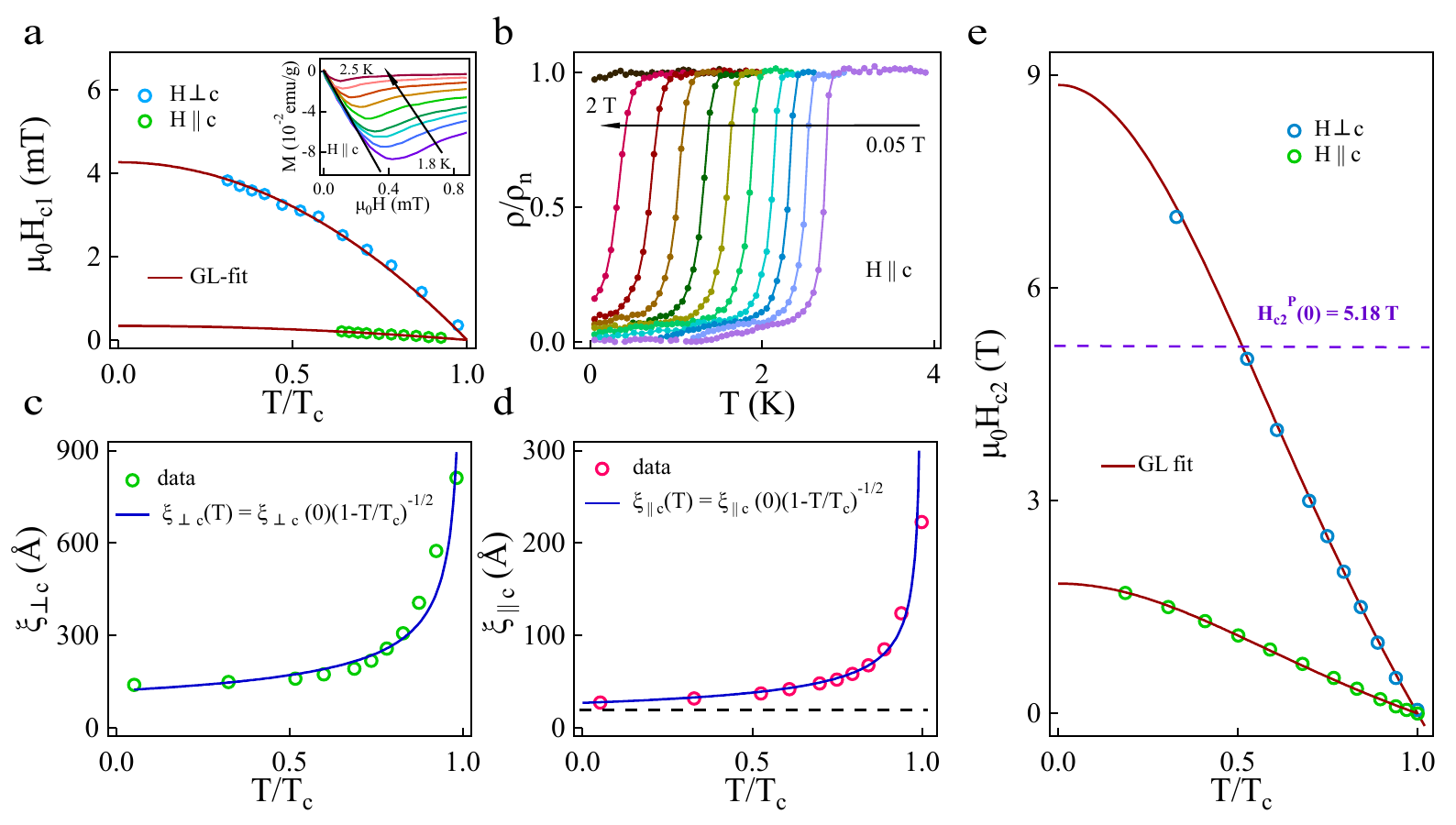}
\caption {\label{fig2} (a) The variation of lower critical field ($H_{c1}$) under different temperatures in two measured directions ($H \parallel c$ and $H \perp c$). The purple solid line is the GL-fitted curve. Inset: Field-dependent magnetization curves in $H \parallel c$ orientation. (b) The normalized temperature-dependent electrical resistivity at different fields ranging from 0.05 T to 2 T in the $H \parallel c$ direction. (c) and (d) The in-plane and out-of-plane coherence lengths vary with reduced temperature ($T/T_c$). The maroon solid curves are the fitted results. The black dashed line in (d) indicates the inter-layer spacing. (e) The temperature-dependent upper critical fields in both directions are well fit by the GL model (red solid line). The violet dashed line shows the Pauli limiting field.}
\end{figure*}

\section{RESULTS AND DISCUSSION}
\subsection{Sample Characterization}
\figref{fig1}(a) shows the crystal structure of 3$R$-Ta$_{0.9}$W$_{0.1}$Se$_2$, generated by VESTA software \cite{momma2011vesta}. The Rietveld refinement \cite{fullprof} of the powder X-ray diffraction patterns at room-temperature of the crushed crystal (shown in \figref{fig1}(b)) confirms the 3$R$-phase in W-doped TaSe$_2$. This compound has a space group $R$ 3 m (160) with a rhombohedral structure. The refined lattice parameters are $a$ = $b$ = 3.422(9), $\text{\AA}$, $c$ = 19.187(1) $\text{\AA}$. W doping is supported by the slight increase in the $c$-value and the corresponding decrease in the $a$-value compared to the parent TaSe$_2$ \cite{bjerkelund1967structural}. The left inset of \figref{fig1}(c) displays the microscopic image of the 3 mm long crystal, while the right inset shows the Laue diffraction pattern, confirming the crystalline nature of the sample. \figref{fig1}(c) presents the single crystal XRD indicating that the orientation of the crystals is along the $(00n)$ direction with the $c$-axis as a growth axis. The elemental composition is further verified using EDAX analysis (see supplemental material \cite{Supp}, Fig. \textcolor{blue}{S1}).

\subsection{Electrical Resistivity and Magnetization}
The zero-field electrical resistivity of the 3$R$-Ta$_{0.9}$ W$_{0.1}$Se$_2$ crystal was measured in the temperature range of 0.05 $-$ 300 K, as reflected in \figref{fig1}(d). The resistivity curve exhibits a CDW transition around 102 K, slightly lower than that of the parent compound, with a residual resistivity ratio (RRR) of roughly 2.5 ($\rho(300 K)$/$\rho(3 K)$). The superconducting transition occurred at 2.82(2)K as shown in the inset of \figref{fig1}(d), where the zoomed view near the superconducting region reveals a relatively broader resistive transition compared to the parent 3$R$-TaSe$_2$ \cite{https://doi.org/10.1002/adfm.202501453}, thus confirming the successful doping of W in 3$R$-TaSe$_2$.

The Hall resistivity measurement was performed using the four-probe method to determine the carrier density. \figref{fig1}(e) shows the linear variation of Hall resistivity under an applied magnetic field at different temperatures from 6 K to 250 K. The slope ($R_H$) of the $\rho_{xy}$ $-$ $H$ curves reveals that its sign changes from positive to negative after $T_{CDW}$, implying a crossover in the dominant charge carriers \cite{https://doi.org/10.1002/adfm.202501453, m9xx-gk46}. By linear fitting the curve $\rho_{xy}$ $-$ $H$ for 6 K, the obtained value of $R_H$ is 9.76(8) $\times$ 10$^{-12}$ cm$^3$C$^{-1}$. The corresponding carrier density $n$ = 6.39(8) $\times$ 10$^{27}$ m$^{-3}$ was calculated using the formula: $R_H$ = 1/$n$e. We have also derived Hall coefficients for other temperatures (Fig. \textcolor{blue}{S2(c)} in supplemental material \cite{Supp}).

Superconductivity is verified by performing temperature-dependent magnetization measurements. Two standard mechanisms: zero-field-cooled warming (ZFCW) and field-cooled cooling (FCC), under 1 \si{mT} were used to determine the superconducting transition. A clear diamagnetic response in both modes confirms the superconducting nature of 3$R$-Ta$_{0.9}$W$_{0.1}$Se$_2$ at 2.82(1) K, as illustrated in \figref{fig1}(f). The relatively weaker magnetization signal in the FCC mode compared to the ZFCW mode is attributed to stronger flux pinning. 

Field-dependent magnetization measurements ($M$) were performed at several temperatures to determine the lower critical field ($H_{c1}$(0)). The $H_{c1}$ values were extracted from the $M$ $-$ $H$ curves at the deviation from the linear Meissner response, indicating the onset of magnetic flux penetration. To identify the anisotropic nature of the superconducting state, these measurements were executed for two field orientations: with the magnetic field applied parallel (inset of \figref{fig2}(a)) and perpendicular (Fig. \textcolor{blue}{S2(a)}, supplemental material \cite{Supp}) to the growth axis $c$, respectively. The extracted $H_{c1}$(T) were subsequently plotted against the reduced temperature ($T/T_c$) and analyzed using the conventional Ginzburg-Landau (GL) framework \cite{tinkham2004introduction}, expressed as
\begin{equation}
H_{c1}(T)=H_{c1}(0)\left[{1-t^{2}}\right], \quad  \text{where} \;  t = \frac{T}{T_{c}}
\label{eqn3:HC1}
\end{equation}
This extrapolates $H_{c1}$(0) as 0.33(7) and 4.27(1) mT for $H$ parallel and perpendicular to the $c$-axis, respectively, as shown in \figref{fig2}(a). To determine the upper critical field ($H_{c2}$(0)), temperature-dependent resistivity measurements were performed under different magnetic fields, as shown in \figref{fig2}(a) and \textcolor{blue}{S2}(b) (supplemental material, \cite{Supp}) for $H$ aligned perpendicular and parallel to the same plane, respectively. The superconducting transition temperature ($T_{c}$) is gradually decreased with increasing magnetic field. We incorporate a widely used criterion for a broader resistive transition, defined at $\rho$ = 0.95$\rho_n$, where $\rho_n$ is the resistivity of the normal-state just above the transition \cite{https://doi.org/10.1002/adfm.202501453, doi:10.1021/jacs.4c09248}. The applied magnetic fields were plotted as a function of $t$ and then examined using the standard GL relation (\equref{eqn4:HC2}) which extrapolates $H_{c2}$(0) as 1.83(2) T for $H \parallel$ c and 8.86(1) T for $H \perp$ c, respectively, as given in \figref{fig2}(e).
\begin{equation}
H_{c2}(T) = H_{c2}(0)\left[\frac{1-t^{2}}{1+t^{2}}\right].
\label{eqn4:HC2}
\end{equation}

\begin{figure}
\centering
\includegraphics[width=0.95\columnwidth]{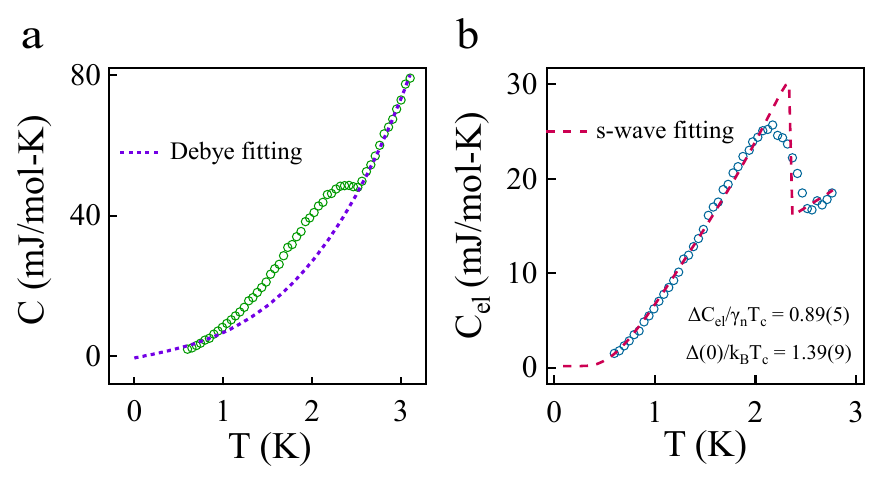}
\caption {\label{fig3} (a) The Debye-Sommerfeld model is used to fit the normal-state specific heat data. (b) $C_{el}$ is fitted as a function of sample temperature using an isotropic s-wave model.}
\end{figure}

The different values of $H_{c2}$(0), along with those of $H_{c1}$(0), give insight into the anisotropic nature of superconductivity, and this anisotropy is quantified by calculating the parameter $\Gamma$ = ${Hc_2^{\parallel c}}/{Hc_2^{\perp c}}$ as 0.20. The pronounced anisotropy in this system suggests a quasi-2D superconducting state. This quasi-2D behavior is further corroborated by angle-dependent upper-critical-field measurements and the observation of a BKT transition.

The suppression of superconductivity by an external magnetic field is mainly governed by two basic mechanisms: the Pauli and orbital limiting effects. Zeeman splitting, which breaks Cooper pairs by aligning electron spins parallel to the field, is the main reason for the Pauli-limiting effect. In contrast, the Lorentz force-driven orbital-limiting effect leads to vortex formation, thereby disrupting the coherent motion of Cooper pairs. In the context of the BCS theory, the Pauli limiting effect is estimated at 5.18 T using $H_{c2}^{P}$(0) = 1.84 $T_{c}$ with $T_{c}$ = 2.82 K \cite{Chandrasekhar1962pauli, Clogston1962pauli}. This value is lower than the experimentally observed in-plane upper critical field, indicating the violation of the Pauli limit by a factor of about 1.7. This pronounced violation of the Pauli limit points to the possible presence of Ising SOC, underscoring the unconventional nature of the superconducting state \cite{zhang2022tailored}. However, a detailed theoretical analysis is needed to identify the origin of the Pauli limit violation in 3$R$-Ta$_{0.9}$W$_{0.1}$Se$_2$.

The upper critical field ($H_{c2}$) and the superconducting coherence length ($\xi$) are quantitatively related through a well-established theoretical framework \cite{palstra1988angular}, described as: 
\begin{equation}
    H_{c2} = \frac{\phi_0}{2\pi\xi_{\perp c}^2}(cos^2\theta+\epsilon^2sin^2\theta)^{-1/2},
    \label{eqn4: coherence}
\end{equation} 
where the magnetic flux quanta, $\phi_{0}$(h/2e) has a value of 2.07$\times$10$^{-15}$ T$m^{2}$. The anisotropy parameter $\epsilon$ is the ratio between the two coherence lengths along the in-plane and out-of-plane directions, denoted as $\epsilon$ = $\xi_{\parallel c}$/$\xi_{\perp c}$. $\theta$ is the angle between the unit vector normal to the superconducting layers and the direction of the applied magnetic field. By limiting the angular dependence of the upper critical field [\equref{eqn4: coherence}] in $\theta = 0^\circ$ and $90^\circ$, simplified expressions are obtained for the coherence lengths in the parallel and perpendicular directions to the applied field. So, the anisotropic GL-equations are expressed as $H_{c2}^{\parallel}(0) = \frac{\phi_0}{2\pi \xi_{\perp}^2}$ and for $H_{c2}^{\perp}(0) = \frac{\phi_0}{2\pi \xi_{\parallel} \xi_{\perp}}$. \figref{fig2}(c) shows that $\xi_{\perp}(T)$ follows the GL prediction over the entire temperature range, obeying the simple relation $\xi(T) = \xi(0)(1 - T/T_c)^{-1/2}$. On the other hand, $\xi_{\parallel}(T)$ (\figref{fig2}(d)) decreases more rapidly than expected from GL theory and becomes comparable to the inter-layer distance near the low temperature region. This behavior has also been reported in other Josephson-coupled superconductors with weak inter-layer coupling \cite{PhysRevLett.44.892, PhysRevB.110.174511}, suggesting that a dimensional crossover can arise due to the significant weakening of the inter-layer coupling. The coherence lengths obtained in both directions are 2.67(9) and 12.0(5) nm, respectively.

\begin{figure*}
\includegraphics[width=2\columnwidth]{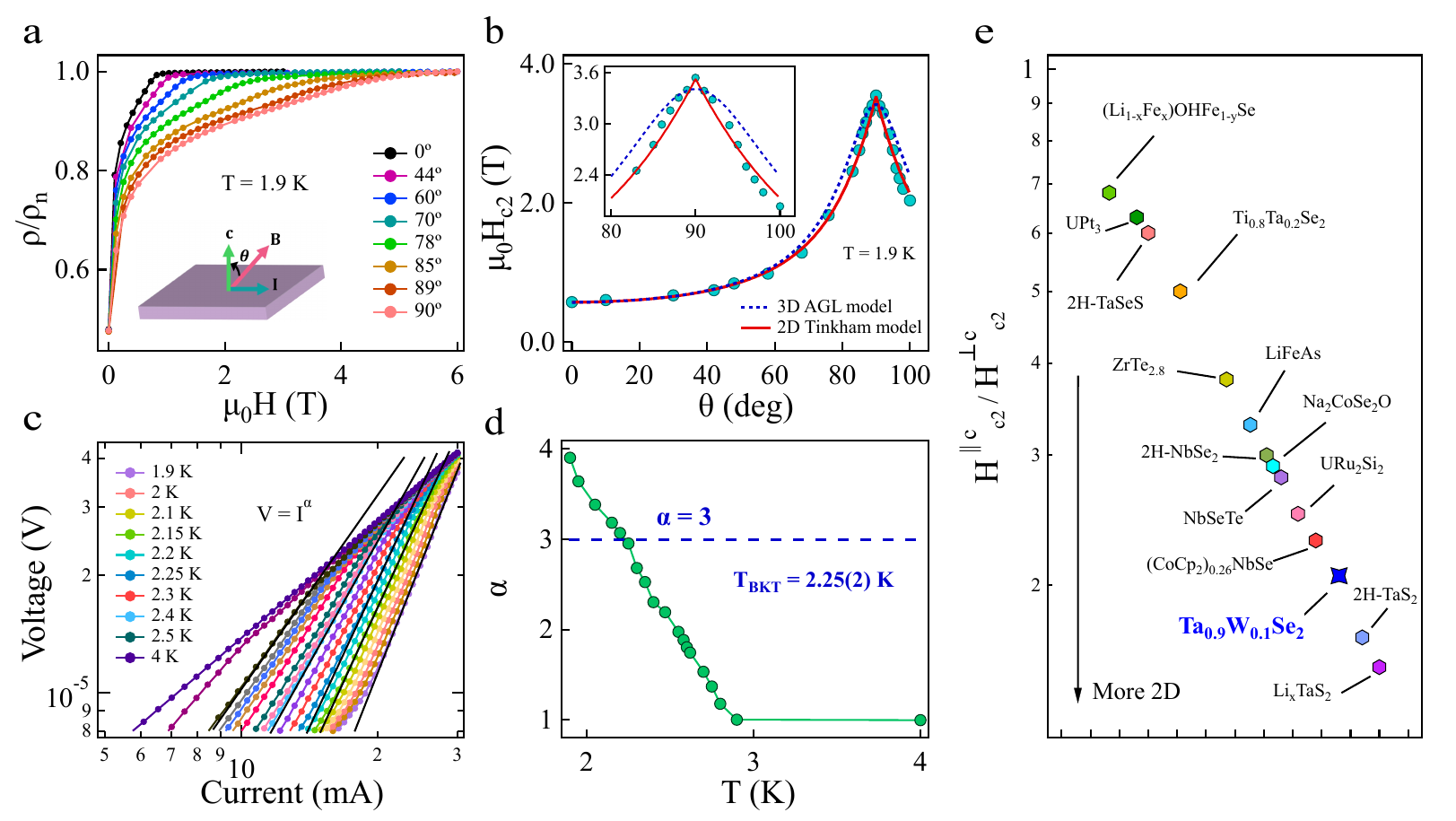}
\caption {\label{fig4}(a) Electrical resistivity versus magnetic field measured at multiple angles shows an anisotropic characteristic. Inset: Schematic illustration of the corresponding experimental setup. (b) The upper critical field as a function of the measured angles is fitted using the 2D Tinkham model and the 3D AGL model. Inset is an enlarged view of the fitting near a 90$^{\circ}$ angle. (c) The temperature-dependent voltage-current ($V$-$I$) curves at zero magnetic field are plotted on a logarithmic scale. (d) The exponent ($\alpha$) values from power-law fitting correspond to the BKT transition temperature at 2.91(1) K. The slope of $\alpha (T)$ = 3 is marked by the horizontal dotted line. (e) The anisotropic parameter for 3$R$-Ta$_{0.9}$W$_{0.1}$Se$_2$ is compared with various superconducting materials ${Hc_2^{\parallel c}}/{Hc_2^{\perp c}}$.}
\end{figure*}

A set of standard theoretical relations is used to determine the GL penetration length ($\lambda$) and the GL parameter ($\kappa$): $H_{c2}^\perp$(0)/$H_{c1}^\perp$(0) = 2$\kappa_{\perp c}^2/ln \kappa_{\perp c}$, $\kappa_{\perp c}$ = [$\lambda_{\perp c}$(0)$\lambda_{\parallel c}$(0)/$\xi_{\perp c}$(0)$\xi_{\parallel c}$(0)]$^{1/2}$, and $\kappa_{\parallel c}$ = $\lambda_{\perp c}$(0)/$\xi_{\perp c}$(0). The extracted superconducting parameters are summarized in \tableref{tbl1}. The obtained values of $\kappa$ for both field orientations are well above the critical limit of $1/\sqrt{2}$, clearly establishing that 3$R$-Ta$_{0.9}$W$_{0.1}$Se$_2$ falls into the class of strong type-II superconductors. Furthermore, the thermodynamic critical field ($H_c$), which measures the superconducting condensation energy, was calculated to be approximately 2 T, using the relation $H_c(0) = H_{c1}^\perp(0)\sqrt{2}\kappa_{\perp c}/ln \kappa_{\perp c}$.

\subsection{Specific Heat}
We further verified the superconducting nature of this compound using specific heat measurements along with resistivity and magnetization data. Polycrystalline 3$R$-Ta$_{0.9}$W$_{0.1}$Se$_2$ shows a distinct jump at 2.5(1) K (\figref{fig3}). Temperature-dependent electronic specific data, fitted with an isotropic s-wave model, reveal weakly coupled single-gap superconductivity (details in the supplemental material \cite{Supp}). More importantly, the calculated density of states (DOS) at the Fermi energy is enhanced compared to other doped and intercalated superconducting TaSe$_2$ systems \cite{wan2023superconducting}. A detailed comparison of key physical parameters on some doped and intercalated TaSe$_2$ is presented in \tableref{tbl2}.

\begin{table}[b]
\caption{The anisotropic superconducting parameters of the synthesized single crystal.}
\label{tbl1}
\setlength{\tabcolsep}{13pt}
\renewcommand{\arraystretch}{1.25}
\begin{center}
\begin{tabular}[b]{lccc}\hline
Parameters& Unit & $H\parallel c$ & $H\perp c$ \\
\hline
$H_{c1}(0)$ & mT & 0.33(7) &   4.27(1) \\ 
$H_{c2}^{res}$(0) & T & 1.83(1) &   8.86(4) \\
$\xi$ & nm & 2.67(9) &  12.0(5)  \\
$\lambda$ & nm & 101.4(7)  & 1385.7(5)  \\
$\kappa$& & 115 & 66  \\
\hline
\end{tabular}
\end{center}
\end{table}

\subsection{Anisotropic Superconductivity}
\textbf{Angle-dependent upper critical field.}
Quasi-two-dimensional superconductivity, along with the identification of an anisotropic nature, is probed using angle-dependent upper critical field measurements. To this end, field-dependent resistivity measurements were performed at a fixed temperature of 1.9 K for various field orientations, as illustrated in \figref{fig4}(a). Here, $\theta$ infers the angle between the applied magnetic field and the normal to the sample plane, while the current is maintained along the $ab$ plane. The upper critical field at every angle was determined using the criterion $\rho = 0.95\rho_n$, where $\rho_n$ represents the resistivity in the normal state. As depicted in \figref{fig4}(b), the angular dependence of $H_{c2}$ (pink symbols) exhibits a pronounced cusp near $\theta = 90^\circ$, which is further highlighted in the inset. This similar feature is mainly associated with quasi-2D superconductivity. To further verify this behavior, two theoretical models are imposed here: the anisotropic Ginzburg–Landau (AGL) model (\equref{eqnAGL}) \cite{https://doi.org/10.1002/adfm.202208761} and the Tinkham model (\equref{eqn2DT}) \cite{PhysRev.129.2413}. The former describes the angular variation in three-dimensional (3D) superconductors, with an ellipsoidal form of $H_{c2}(\theta)$, whereas the latter accounts for two-dimensional (2D) thin-film superconductors and predicts a characteristic cusp at $\theta = 90^\circ$. Therefore, fitting the experimental data to these models allows us to evaluate the dimensionality of the superconducting state in this system.
\begin{equation}\label{eqnAGL}
    \left(\frac{H_{c2}(\theta,T) \sin{\theta}}{H_{c2}^{\perp}}\right)^2 + \left(\frac{H_{c2}( \theta,T) \cos{\theta}}{H_{c2}^{||}}\right)^2 = 1
\end{equation}

\begin{equation}\label{eqn2DT}
    \left(\frac{H_{c2}(\theta,T) \sin{\theta}}{H_{c2}^{\perp}}\right)^2 + \left\arrowvert\frac{H_{c2}( \theta,T) \cos{\theta}}{H_{c2}^{||}}\right\arrowvert = 1
\end{equation}
The analysis is more accurately explained by the 2D Tinkham model than by the 3D AGL model \cite{patra2024planar}, suggesting a significant quasi-2D behavior in superconducting 3$R$-Ta$_{0.9}$W$_{0.1}$Se$_2$.\\

\begin{figure*}
\includegraphics[width=2\columnwidth]{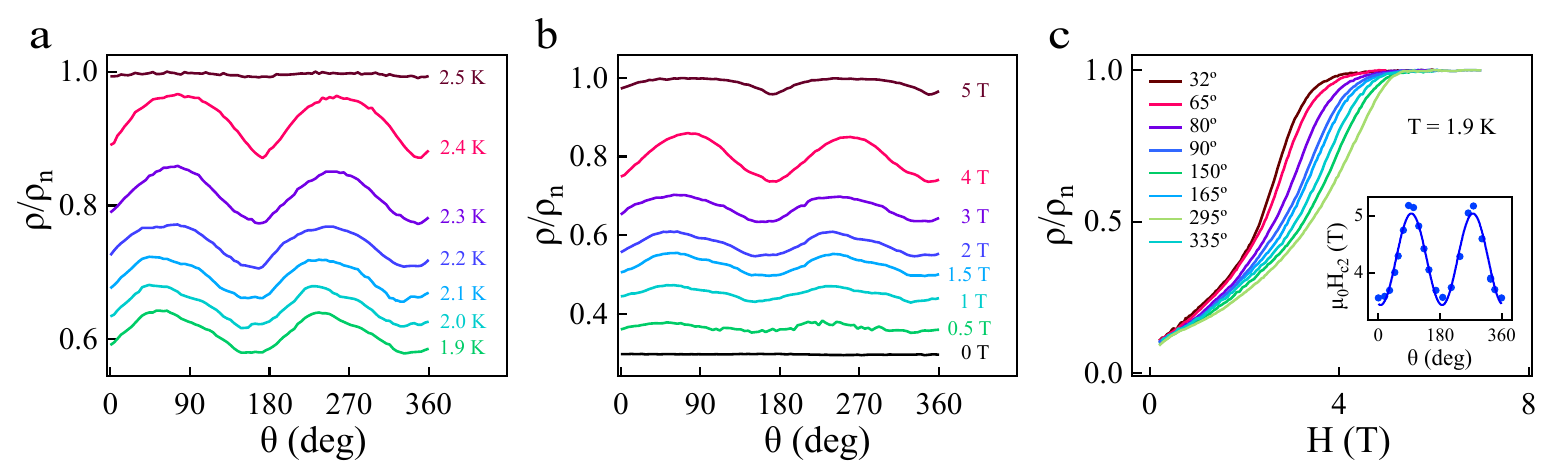}
\caption {\label{fig5} (a) and (b) Angle-dependent normalized resistivity under various temperatures and magnetic fields in the superconducting region, respectively. (c) In-plane angular dependence of resistivity measured under an applied magnetic field at T = 1.9 K. Inset: The angular variation of in-plane upper critical field, exhibiting a pronounced two-fold oscillatory behavior.}
\end{figure*}

\textbf{Berezinskii-Kosterlitz-Thouless (BKT) transition.}
Along with angle-dependent transport measurements, identifying a Berezinskii-Kosterlitz-Thouless (BKT) transition is another useful method to probe the quasi-2D behavior of layered superconductors. 2D superconductors can exhibit quasi-long-range correlations due to increased thermal fluctuations, while 3D superconductors allow for actual long-range phase coherence. When a system evolves from a quasi-long-range ordered vortex–antivortex phase to a fully disordered state with increasing temperature \cite{takiguchi2024berezinskii},  a BKT transition emerges near the superconducting transition. As shown in \figref{fig4}(c), the temperature-dependent current-voltage ($V$-$I$) curves are analyzed to determine the BKT transition temperature. At temperatures well below $T_c$, the presence of sufficiently high currents can unbind vortex-antivortex pairs, leading to a non-linear behavior (\equref{eqnV-I}) that follows a power-law dependence. Here, the exponent ($\alpha$) is directly proportional to the superfluid density ($J_s$) \cite{PhysRevB.100.064506}. 
\begin{equation}
    V \propto I^{\alpha(T)},  
    \quad \alpha(T) = 1 + \pi \frac{J_s(T)}{T} .
    \label{eqnV-I}
\end{equation}

At $T_{BKT}$, $J_s$ attains the value 2/$\pi T_{BKT}$, resulting in $\alpha$ = 3. Accordingly, by fitting the $V$-$I$ curves, we have calculated $\alpha$ and plotted it as a function of temperature, as shown in \figref{fig4}(d). From this analysis, the transition temperature of BKT is estimated to be 2.25 K, which corresponds to $\alpha$ = 3. Above $T_c$, the $V$-$I$ curves retain ohmic-behavior with $\alpha$ = 1. The observation of a BKT transition thus implies that 3$R$-Ta$_{0.9}$W$_{0.1}$Se$_2$ behaves like a quasi-2D superconductor.

The dimensionality of superconductivity can be qualitatively inferred from the anisotropy of the upper critical field, with the ratio ${Hc_2^{\parallel c}}/{Hc_2^{\perp c}}$ decreasing as the system evolves from a 3D to a quasi-2D regime. A comparative overview of the anisotropy parameters in bulk and monolayer superconductors is presented in \figref{fig4}(e).\\

\begin{table}[b]
\caption{Summary of key physical parameters for multiple superconducting doped and intercalated TaSe$_2$ compounds.}
\label{tbl2}
\setlength{\tabcolsep}{2.8pt}
\renewcommand{\arraystretch}{1.25}
\begin{center}
\begin{tabular}[b]{lcccc}\hline
Compound & $T_c$ (K) & $\lambda_{ep}$ & \makecell{$N(E_F)$ \\ states/eV/f.u.} & Ref. \\
\hline
2$H$-TaSe$_2$ & 0.14 & 0.4 & 1.51 & \cite{bhoi2016interplay}\\
2$H$-Pd$_{0.09}$TaSe$_2$ & 3.3 & 0.67 & 2.16 & \cite{bhoi2016interplay}\\
3$R$-Ta$_{0.9}$W$_{0.1}$Se$_2$ & 2.82 & 0.63 & 2.76 & This work\\ 
3$R$-TaSe$_{2-x}$Te$_x$ & 2.4 & 0.64 & 1.88 & \cite{doi:10.1073/pnas.1502460112} \\
\hline
\end{tabular}
\end{center}
\end{table}

\textbf{Anisotropic magnetoresistance in the superconducting state.} We have further investigated the in-plane anisotropy of the upper critical field of 3$R$-Ta$_{0.9}$W$_{0.1}$Se$_2$. Under in-plane rotation of magnetic fields, magneto-resistance (MR) shows a pronounced two-fold modulation in the superconducting state, as shown in \figref{fig5}(a) and (b), respectively. Panel (a) displays angle-dependent MR at various temperatures under a fixed field of 0.5 T, while panel (b) shows angle-dependent MR under varying fields at 1.9 K. Field-dependent resistivity measurements were performed at 1.9 K to examine the variation of the in-plane upper critical fields with angle (\figref{fig5}(c)). The extracted upper critical field is plotted as a function of angle, which follows a sinusoidal behavior of the form cos(2 $\theta$ + $\phi$) (here, $\phi$ represents the phase relative $\theta = 0^\circ$), as represented by the solid blue line in the inset of \figref{fig5}(c). Notably, although 3$R$-TaSe$_2$ intrinsically possesses a three-fold rotational symmetry, the superconducting state reveals a significant two-fold anisotropy \cite{https://doi.org/10.1002/adfm.202501453}. Together with the Pauli limit violation of the in-plane upper critical field and noncentrosymmetric crystal structure, this behavior points to an underlying symmetry-breaking field. This field can facilitate mixing between the dominant s-wave pairing with subdominant p- or d-wave channels, suggesting an unconventional superconducting state \cite{hamill2021two}.

The angle-dependent upper critical field and the BKT transition confirm the quasi-2D nature of 3$R$-Ta$_{0.9}$W$_{0.1}$Se$_2$ \cite{PhysRevMaterials.4.124803}. This behavior suggests that substituting Ta with W weakens interlayer coupling and increases the density of states at the Fermi energy. In the monolayer limit, the superconductivity is typically stabilized against high magnetic fields by intrinsic Ising SOC. As in the monolayer counterpart, our bulk sample exhibits an in-plane upper critical field that exceeds the Pauli limit, suggesting unconventional Ising superconductivity. Additionally, the observed rotational symmetry breaking in the superconducting state provides evidence for unconventional pairing. These results prompt the need for a more detailed theoretical understanding of the exact superconducting pairing mechanism in 3$R$-Ta$_{0.9}$W$_{0.1}$Se$_2$.

\section{Conclusion}
We have performed magnetization, specific heat and transport measurements on CVT-grown single crystals of 3$R$-Ta$_{0.9}$W$_{0.1}$Se$_2$. Together, the measurements reveal a superconducting transition at 2.82(2) K, which confirms this compound as a type-II superconductor. Remarkably, the in-plane upper critical field exceeds the Pauli limit by a factor of 1.7, suggesting the possible emergence of Ising pairing and highlighting the unconventional nature of the superconductivity. Specific heat measurements also imply a weakly coupled superconducting nature. The angular dependence of the upper critical field, well described by the 2D Tinkham model, and the occurrence of the BKT transition at 2.25 K, substantiate the quasi-2D nature of this sample. Furthermore, the in-plane magnetoresistance in the superconducting state exhibits a clear rotational symmetry breaking, constituting a striking signature of unconventional superconductivity. The microscopic origin of this anisotropic behavior remains unresolved and may have several possible explanations. While this nature may imply an intrinsic nematic pairing or coupling between superconductivity and CDW order, it may also originate due to an externally applied symmetry breaking field that can induce a two-fold response with Ising pairing. Although the combination of Pauli limit violation and field-induced mixed-parity states hints at unconventional superconducting pairing, additional thickness-dependent measurements are required to better understand them. Additionally, the easy exfoliation nature of this material provides a promising route to explore diverse low-dimensional quantum phases in bulk noncentrosymmetric TMDs, considerably broadening a new pathway for realizing 2D superconductivity with unconventional pairing mechanisms.

\section{METHODS}
\noindent\textit{Experimental Details.} Single crystals of 3$R$-Ta$_{0.9}$W$_{0.1}$Se$_2$ were grown for the first time using the conventional chemical vapor transport (CVT) method, with iodine (I$_2$) as the transport agent. High-purity powders of Ta (5N), W (4N) and Se (5N) were mixed in appropriate stoichiometric ratios and sealed in an evacuated quartz ampule together with I$_2$. The ampule was then placed in a three-zone tubular furnace and subjected to a temperature gradient of 100 K, with a hot-zone temperature of 1093 K. After a 14-day reaction, the furnace was allowed to cool naturally, producing shiny black crystals. Powder X-ray diffraction (PXRD) on an X'pert PANalytical Empyrean X-ray diffractometer with monochromatic Cu- $K_\alpha$ radiation ($\lambda$ = 1.54 \AA) was used to analyze the identity and phase purity of the compound. The crystalline nature was further verified by a Laue diffraction pattern obtained with a Photonic Science Laue camera. Elemental composition was confirmed using energy-dispersive X-ray analysis (EDAX) along with scanning electron microscopy (SEM). A Quantum Design magnetic property measuring system (MPMS3) with a $^4He$ cryostat was used to measure magnetic properties. A 9T-Quantum Design physical property measurement system (PPMS) and a dilution refrigerator (DR) were used to perform transport and specific heat measurements using standard four-probe and two-tau methods, respectively.

\section*{Supporting Information}
\noindent Supporting Information is available from the Wiley Online Library or from the author.

\section*{Acknowledgments} 
\noindent P.~M. acknowledges the funding agency DST-INSPIRE, Government of India, for providing the SRF fellowship. R.~P.~S. acknowledges the Science and Engineering Research Board, Government of India, for the Core Research Grant: CRG/2023/000817.

\section*{Conflict of Interest}
\noindent The authors declare no conflict of interest.

\section*{Data Availability Statement}
\noindent Data supporting the findings of this study are available from the corresponding author on a reasonable request.

\nocite{*}
\bibliographystyle{MSP}
\bibliography{Library}
\end{document}